\documentclass[aps,amsmath,nofootinbib,11pt,superscriptaddress]{revtex4}

\begin{document}

\title{Signatures of a minimal length scale in high precision experiments}

\author{U.~Harbach}
\email{harbach@th.physik.uni-frankfurt.de}
\affiliation{
Institut f\"ur Theoretische Physik\\
J. W. Goethe-Universit\"at\\
Robert-Mayer-Str. 8-10\\
60054 Frankfurt am Main, Germany}
\author{S.~Hossenfelder}
\affiliation{
Department of Physics\\
University of Arizona\\
1118 East 4th Street\\
Tucson, AZ 85721, USA}
\author{M.~Bleicher}
\author{H.~St\"ocker}
\affiliation{
Institut f\"ur Theoretische Physik\\
J. W. Goethe-Universit\"at\\
Robert-Mayer-Str. 8-10\\
60054 Frankfurt am Main, Germany}

\begin{abstract}
We discuss modifications of the gyromagnetic moment of electrons and muons due to a minimal length
scale combined with a modified fundamental scale $M_f$. First-order deviations from the
theoretical standard model value for $g-2$ due to these String Theory-motivated effects are derived.
Constraints for the new fundamental scale $M_f$ are given.
\end{abstract}

\maketitle

\section{Motivation and Introduction}

Although the standard model is a powerful tool to explain the physics of the very basic
constituents of matter, it is far from being an exhaustive description of our world. Many
questions remain unanswered: What causes the existence of three particle generations? Where
do the various quark and lepton masses and coupling constants come from? How to unite gravity and
quantum theory? Why is gravity so weak compared to the other forces? Theories such as M-Theory and Superstrings try to
give a hint on these questions, but they do not (yet) provide us with measurable
quantities. Nevertheless, there are some general features that seem to go hand in hand
with all promising candidates for a theory of quantum gravity:

\begin{itemize}
\item the need for a higher dimensional space-time and
\item the existence of a minimal length scale.
\end{itemize}

In this paper, we study implications of these extensions in the Dirac equation
without the aim to derive them from a first principle theory. Instead we will
analyse possible observable modifications that may arise by combining the main features
of both extra dimensions and a minimal length scale in a simplified model.

\section{Large eXtra Dimensions}

The idea of Large eXtra Dimensions (LXDs) which was recently proposed in
\cite{Arkani-Hamed:1998rs,Antoniadis:1998ig,Arkani-Hamed:1998nn,Randall:1999vf,Randall:1999ee}
might allow to study first effects of unification or quantum gravity in near future
experiments. In these models, only gravitons can propagate into the $d$ compactified
LXDs. The standard model particles are bound to our (3+1)-dimensional sub-manifold, often
called our 3-brane. This results in a decrease of the Planck scale to a new fundamental
scale $M_f$ and gives rise to the exciting possibility of TeV scale GUTs
\cite{Dienes:1998qh}. Therefore, not only the notion of further dimensions of space-time is
incorporated, but also the hierarchy-problem is solved, although one might claim it is only
shifted to the geometrical sector.

In \cite{Arkani-Hamed:1998rs}, the following relation between the
four-dimensional Planck mass $m_p$ and the higher dimensional fundamental scale $M_f$ is derived:

\begin{eqnarray}\label{fundi}
m_p^2 = R^d M_f^{d+2} \quad,
\end{eqnarray}

where $R$ is the radius of the LXDs. This is a consequence of Gauss' law in $3+d$ spatial dimensions:
Two test masses $m_1$, $m_2$ within a distance below the compactification radius will feel
the gravitational potential

$$\frac{V(r)}{m_1} \sim \frac{1}{M_f^{d+2} r^d} \frac{m_2}{r} \quad,(r \ll R).$$

At distances above the compactification radius, the gravitational flux lines are not further dissolved into the
extra dimensions, and one has to regain the usual potential in three spatial dimensions

$$\frac{1}{M_f^{d+2} R^d} \frac{m_2}{r} \stackrel{r \gg R}{\sim} \frac{1}{m_p^2} \frac{m_2}{r} \quad,$$

which directly yields \eqref{fundi}.

This lowered fundamental scale leads to a vast number of observable phenomena of
quantum gravity at energies in the range of $M_f$. In fact, the non-observation of these predicted features in past
experiments gives first constraints on the parameters
of the model, the number of extra dimensions $d$ and the fundamental scale $M_f$
\cite{Uehara:2002yv,Cheung:1999fj}.
This scenario has major consequences:

\begin{itemize}
\item Cosmology and astrophysics: Modification of inflation in the early universe and enhanced
supernova-cooling due to graviton emission
 \cite{Arkani-Hamed:1998nn,Cullen:1999hc,Hewett:2002hv,Barger:1999jf,Hanhart:2001fx}.
\item Additional processes are expected in high-energetic lepton and hadron interactions
\cite{Mirabelli:1998rt,Gleisberg:2003ue}: production of real and virtual gravitons
\cite{Giudice:1998ck,Giudice:2000av,Hewett:1998sn,Nussinov:1998jt,Rizzo:1999pc} and the
creation of black holes at energies that can be achieved at colliders in the near future
\cite{Argyres:1998qn,Giddings:2001ih,Mocioiu:2003gi,Kotwal:2002wg,Uehara:2001yk,Emparan:2001kf,Hossenfelder:2001dn}
and in ultra high energetic cosmic rays \cite{Ringwald:2001vk}.
\item One also has to expect the influence of the extra dimensions on high precision measurements; the
most obvious being the modification of Newton's law at small distances
\cite{Hoyle:2000cv,Long:1998dk,Chiaverini:2002cb}.
\item Of highest interest are also modifications of the
gyromagnetic moment of Dirac particles which promises new insight into non-standard model
couplings and effects
\cite{Agashe:2001ra,Cacciapaglia:2001pa,Park:2001xp,Appelquist:2001jz,Nath:2001wf,Calmet:2001si}.
\end{itemize}

Thus, new phenomena might either be encountered in high energy or high precision experiments.

\section{The Minimal Scale}

As discussed above, String theory suggests the existence of a minimal length scale. In perturbative string
theory \cite{Gross:1988ar,Amati:1989tn}, the feature of a fundamental minimal length scale arises
from the fact that strings can not probe distances smaller than the string scale. If the
energy of a string reaches the Planck mass $m_p$, excitations of the string can
occur and increase the extension \cite{witten}. Due to this, uncertainty in position
measurement can never become smaller than $l_p = \hbar / m_p$. For a detailed review, the reader is referred to Refs. 
\cite{Garay:1995en,Kempf:1998gk}.

However, in the present model with LXDs, this fact grows important for collider physics at high energies or for high
precision measurements at low energies due to the lowered fundamental scale $M_f$, which results in a new fundamental length scale $L_f = \hbar / M_f$.

Naturally, this minimum length uncertainty is related to a modification of the
standard commutation relations between position and momentum
\cite{Kempf:1995su,Kempf:1997nk}. Application of this is of high interest for
quantum fluctuations in the early universe and inflation
\cite{Hassan:2002qk,Danielsson:2002kx,Shankaranarayanan:2002ax,Mersini:2001su,Kempf:2000ac,Kempf:2001fa,Martin:2000xs,Easther:2001fz,Brandenberger:2000wr}.  We will follow the propositions made in
\cite{Hossenfelder:2003jz,Harbach:2003qz}.

\section{Incorporation into Quantum Theory}

In order to implement the notion of a minimal length $L_f$, let us now suppose
that one increases the momentum $p$ of a particle arbitrarily, but that the wave number
$k$ has an upper bound. This effect leads to pronounced deviations from the linear dependence when $p$ approaches
the scale $M_f$. The physical interpretation of this is that particles can not
possess arbitrarily small Compton wavelengths $\lambda = 2\pi/k$ so that
arbitrarily small scales cannot be resolved anymore.

To incorporate this behaviour, we assume a relation $k=k(p)$ between $p$ and $k$
which is an uneven function (because of parity) and which asymptotically
approaches $1/L_f$. Furthermore, we demand the functional relation between the
energy $E$ and the frequency $\omega$ to be the same as that between the wave
vector $k$ and the momentum $p$. A possible choice for the relations is

\begin{eqnarray}
L_f k(p) &=& \tanh^{1/\gamma} \left[ \left( \frac{p}{M_f} 
\right)^{\gamma} \right] \quad, \label{eq1}\\
L_f \omega(E) &=& \tanh^{1/\gamma} \left[ \left( \frac{E}{M_f} \right)^{\gamma} 
\right]\quad,\label{eq2}
\end{eqnarray}
with a real, positive constant $\gamma$.

In the following, we restrict our study to the low momentum approximation, namely the regime of first
effects including the orders $(p/M_f)^3$. For this purpose, we expand the function
in a Taylor series for small arguments.
 
Because the exact functional dependence is unknown, we assume an arbitrary
factor $\alpha$ in front of the order $(p/M_f)^3$-term. Therefore the
relations for $k(p)$ and $\omega(E)$ which are used in the following
are
\begin{eqnarray}
L_f k(p) &\approx& \frac{p}{M_f} - \alpha \left(\frac{p}{M_f}\right)^{3}
 \label{app1a}\quad,\\
L_f \omega(E) &\approx& \frac{E}{M_f} - \alpha \left( \frac{E}{M_f}
 \right)^{3}\quad,\\
\frac{1}{M_f} p(k) &\approx& k L_f +\alpha\left( k L_f \right)^3 \quad,\\
\frac{1}{M_f} E(\omega) &\approx& \omega L_f+\alpha\left( \omega L_f \right)^3
 \label{app4a} \quad,
\end{eqnarray}
with $\alpha$ being of order one (e.g. $\alpha=1/3$ for $\gamma=1$), but in general negative values of $\alpha$ can not
 be excluded.

This yields to 3$^{\rm rd}$ order
\begin{eqnarray}
\frac{1}{\hbar} \frac{\partial p}{\partial k} &\approx& 1 +
 3\alpha\left(\frac{p}{M_f} \right)^2 \label{diff2a}\quad.
\end{eqnarray}

The quantisation of these relations is straight forward. The commutators between
 $\hat{k}$ and $\hat{x}$ remain in the standard form:
\begin{eqnarray}
[\hat{x_i},\hat{k_j}]= {\rm i}\delta_{ij}\quad.
\end{eqnarray}

Inserting the functional relation between the wave vector and the momentum then
yields the modified commutator for the momentum. With the commutator relation
\begin{eqnarray}
[\,\hat{x}, \hat{A}(k)] = + {\rm i} \frac{\partial A}{\partial k} \quad,
\end{eqnarray}
the modified commutator algebra now reads
\begin{eqnarray}
[\,\hat{x},\hat{p}]= + {\rm i} \frac{\partial p}{\partial k} \quad.
\end{eqnarray}
This results in the generalised uncertainty relation
\begin{eqnarray}
\Delta p \Delta x \geq \frac{1}{2}  \Bigg| \left\langle \frac{\partial
 p}{\partial k} \right\rangle \Bigg| \quad.
\end{eqnarray}

With the approximations (\ref{app1a})-(\ref{app4a}), the results of Ref.
\cite{Hassan:2002qk} are reproduced up to the factor $\alpha$:
\begin{eqnarray}
[\hat{x},\hat{p}] \approx {\rm i} \hbar\left( 1 + 3\alpha\frac{
 \hat{p}^2}{M_f^2} \right)
\end{eqnarray}
giving the generalised uncertainty relation
\begin{eqnarray}
\Delta p \Delta x \geq \frac{1}{2} \hbar \left( 1+ 3\alpha\frac{\langle
 \hat{p}^2 \rangle }{M_f^2} \right) \quad.
\end{eqnarray}

We give the operators in the position representation which is suited best for this purpose:
\begin{eqnarray}
\hat{x} &=& x\quad,\quad \hat{k}= - {\rm i} \partial_x\nonumber\\
\hat{p} &=& \hat{p}(\hat{k}) \quad,  
\end{eqnarray}
yielding the new momentum operator
\begin{eqnarray}
\hat{p}(\hat{k}) \approx - {\rm i} \hbar \left( 1 - \alpha L_f^2 \partial_x^2
 \right)\partial_x \quad.\label{pofk}
\end{eqnarray} 

In ordinary relativistic quantum mechanics the Hamiltonian of the Dirac Particle
 is \footnote{Greek indices run from 0 to 3, roman indices run from 1 to 3.}
\begin{eqnarray}
\hat{H}= {\rm i} \hbar  \partial_0 = \gamma^0\left( {\rm i}
 \hbar\gamma^i\partial_i +m \right).
\end{eqnarray}

This leads to the Dirac Equation
\begin{eqnarray}
\label{dirac} (p\hspace{-1.5mm}/-m)\psi = 0 \quad,
\end{eqnarray}
with the following standard abbreviation $\gamma^\nu A_{\nu} :=
 A\hspace{-1.5mm}/$ and $p_{\nu} = {\rm i} \hbar \partial_{\nu}$. To include the
 modifications due to the generalised uncertainty principle, we start with the
 relation
\begin{eqnarray}
\hat{E}(\omega)= \gamma^0\left(\gamma^i\hat{p}_i(k) + m \right).
\end{eqnarray}
Including the altered momentum wave vector relation $\hat{p}(\hat k)$ from Eq.
 \eqref{pofk}, this yields again Eq. (\ref{dirac}) with the modified momentum
 operator
\begin{eqnarray}
(p\hspace{-1.5mm}/(\hat k)-m)\psi = 0 \quad.
\end{eqnarray}

This equation is Lorentz invariant by construction. It contains in
 position representation 3rd order derivatives in space coordinates and 3rd order time-derivatives. In our approximation, we can solve the
 equation for a single order time derivative by using the energy condition
 $E^2=p^2+m^2$. This leads effectively to a replacement of time derivatives by
 space derivatives:
\begin{eqnarray}
\hbar \hat{\omega} \approx \hat{E} - \alpha\hat{E}^3/M_f^2 =
 \hat{E}\left(1-\alpha\frac{\hat{p}^i\hat{p}_i + m^2}{M_f^2} \right) \; .
\end{eqnarray} 
Inserting the modified $\hat{E}(\omega)$ and $\hat{p}(k)$ and keeping only
 terms up to 3rd order, we obtain the following expression of the Dirac
 Equation:
\begin{eqnarray}
\label{dirac2}
\omega \vert \psi \rangle \approx  \gamma^0\left(\gamma^i\hat{k}_i
 +\frac{m}{\hbar} \right) \left(1-\alpha\frac{\hbar^2 \hat{k}^i\hat{k}_i +
 m^2}{M_f^2} \right) \vert \psi \rangle \; .
\end{eqnarray} 

\section{The Gyromagnetic Moment}

The task is now to derive the modifications of the anomalous gyromagnetic moment
due to the existence of a minimal length. Therefore we assume as usual the
particle is placed inside a homogeneous and static magnetic field $B$. Regarding the
energy levels of an electron the magnetic field leads to a splitting of the 
energetic degenerated values which is proportional to the magnetic field $B$ and
the gyromagnetic moment $g$. Since the energy of the particle in the field is
not modified (see (\ref{dirac})) there is no modification of the splitting as one
might have expected from the fact that the particles spin is responsible for the
anomaly.

However, if we look at the precession of a dipole in a magnetic field without
minimal length and compare its precession frequency to that of the spin $1/2$ particle
under investigation, again the factor $g$ occurs. Without minimal length the frequency
from quantum mechanics is two times the classical one. In that case a further
modification from the minimal length is expected, as has been under investigation in an alternative
approach in \cite{Schuller:2002rr}. In our model, this modification results from the new
relation between energy and frequency.

Equation (\ref{dirac2}) with minimally coupled electromagnetic fields reads:
\begin{eqnarray}
\omega \vert\psi\rangle \approx \gamma^0\left( \gamma^i\hat{K}_i
 +\frac{m}{\hbar}\right)\left(1-\alpha\frac{\hbar^2 \hat{K}^i\hat{K}_i +
 m^2}{M_f^2}\right)\vert\psi\rangle
\end{eqnarray} 
where $\hat{K}=\hat{k}+ e \hat{A}/\hbar$. Higher derivatives acting on the
magnetic potential can be dropped too for a static and uniform field. In
 addition, the
constant electric potential can be set to zero. In the non-relativistic
 approximation 
we can simplify this equation in the Coulomb gauge to:
\begin{eqnarray}
\left( E+m\hat{F}  \right) \vert \chi\rangle =
\left( \frac{(\hbar\hat{K})^2}{2 m}\hat{F}+\frac{e\hbar}{2
 m}\sigma\hat{B}\hat{F}  \right) \vert\chi\rangle
\end{eqnarray} 
with 
\begin{eqnarray}
\hat{F}=\left( 1- \alpha \frac{\hbar^2 \hat{K}^i\hat{K}_i+m^2}{M_f^2} \right)
 \quad,\quad \vert\psi\rangle = \left\vert{\chi\atop\phi}\right\rangle \quad.
\end{eqnarray} 
Here $\chi$ is the upper component of the Dirac spinor and $\sigma$ denotes the
 Pauli matrices. 

Therefore, the modified expression $\tilde{g}$ for the gyromagnetic moment for $k\rightarrow 0$ is:
\begin{eqnarray}
\tilde{g}=g\cdot\bigg(1-\alpha \frac{m^2}{M_f^2}\bigg)\quad.
\end{eqnarray} 

The experimental data concerning the muon gyromagnetic moment are as follows:
 Davier and collaborates provide two standard model theory results; they differ
 in the experimental input\footnote{The indices indicate the source of the
 vector spectral functions; they are obtained by either hadronic $\tau$ decays
 or $e^+e^-$-annihilation cross-sections.} used to the hadronic contributions
 \cite{Davier:2003pw}. It is convenient to use the quantity $a_\mu = (g-2)/2$ to denote the gyromagnetic factor of the muon:
\begin{eqnarray}
a_{\mu,\tau} &=& 11659195.6(11.1) \times 10^{-10}\nonumber\\
a_{\mu,e^+e^-} &=& 11659180.9(9.7) \times 10^{-10}\quad.\nonumber
\end{eqnarray}

The experimental 'world average' is \cite{Bennett:2002jb}:
\begin{equation}
a_\mu=11659203(8) \times 10^{-10}\quad.
\end{equation}

The results indicate that modifications to the standard model calculation have
 to
be smaller than $10^{-8}$. This leads to the following constraint on the
 fundamental scale
of the theory:
\begin{equation}
M_f/\sqrt{\vert\alpha\vert} \ge 1~{\rm TeV}\quad.
\end{equation}

For the commonly used setting $\gamma=1$ ($\alpha=1/3$), a specific limit on the fundamental
scale $M_f$ can be obtained from present $g-2$ data: $M_f \ge 577$~GeV.

Note that there might
 further be corrections due to graviton loops \cite{Graesser:1999yg,Kim:2001rc}.
 However, recent calculations show that neither sign nor value of these
 corrections are predictable due to unknown form-factors and cutoff parameters
 \cite{Contino:2001nj}.

\section{Summary}

A phenomenological model, which combines both Large Extra Dimensions and the minimal length scale
$L_f$ is studied. The existence of a minimal length scale leads to modifications of quantum
mechanics. With the recently proposed idea of Large Extra Dimensions, this new scale might
be in reach of present day experiments. The modified Dirac equation is used to derive
first-order deviations of the gyromagnetic moment of spin $1/2$ particles. Our results
for the muon $g-2$ value are compared to the values predicted by QED and experiment.

\section*{Acknowledgements}

The authors thank  S.~Hofmann and J.~Ruppert for fruitful discussions. 
This work was supported by the Graduiertenkolleg (Land Hessen), DFG, GSI and
 BMBF.

\bibliography{bormio}
\bibliographystyle{apsrev}

\end{document}